\documentclass[sigconf]{acmart}
\usepackage[ruled,vlined]{algorithm2e}
\usepackage{enumitem}
\usepackage{subfigure}
\usepackage{bm}
\usepackage{enumitem}
\usepackage{marvosym}
\AtBeginDocument{%
  \providecommand\BibTeX{{%
    \normalfont B\kern-0.5em{\scshape i\kern-0.25em b}\kern-0.8em\TeX}}}

\setcopyright{acmcopyright}
\copyrightyear{2018}
\acmYear{2018}
\acmDOI{XXXXXXX.XXXXXXX}

\acmConference[Conference acronym 'XX]{Make sure to enter the correct
  conference title from your rights confirmation emai}{June 03--05,
  2018}{Woodstock, NY}
\acmPrice{15.00}
\acmISBN{978-1-4503-XXXX-X/18/06}

\copyrightyear{2023}
\acmYear{2023}
\setcopyright{acmlicensed}\acmConference[CIKM '23]{Proceedings of the 32nd ACM International Conference on Information and Knowledge Management}{October 21--25, 2023}{Birmingham, United Kingdom}
\acmBooktitle{Proceedings of the 32nd ACM International Conference on Information and Knowledge Management (CIKM '23), October 21--25, 2023, Birmingham, United Kingdom}
\acmPrice{15.00}
\acmDOI{10.1145/3583780.3615456}
\acmISBN{979-8-4007-0124-5/23/10}

\begin{document}

\title{An Incremental Update Framework for Online Recommenders with Data-Driven Prior}
\settopmatter{authorsperrow=5}
\author{Chen Yang}
\affiliation{%
  \country{JD.com}
}
\email{yangchen198@jd.com}

\author{Jin Chen\textsuperscript{\Letter}}
\affiliation{%
  \country{UESTC}
}
\email{chenjin.uestc@gmail.com}

\author{Qian Yu}
\affiliation{%
  \country{JD.com}
}
\email{yuqian81@jd.com}

\author{Xiangdong Wu}
\affiliation{%
  \country{JD.com}
}
\email{wuxiangdong5@jd.com}

\author{Kui Ma}
\affiliation{%
  \country{JD.com}
}
\email{makui@jd.com}

\author{Zihao Zhao}
\affiliation{%
  \country{JD.com}
}
\email{zhaozihao3@jd.com}

\author{Zhiwei Fang}
\affiliation{%
  \country{JD.com}
}
\email{fangzhiwei2@jd.com}

\author{Wenlong Chen}
\affiliation{%
  \country{JD.com}
}
\email{chenwenlong17@jd.com}

\author{Chaosheng Fan}
\affiliation{%
  \country{JD.com}
}
\email{fanchaosheng1@jd.com}

\author{Jie He}
\affiliation{%
  \country{JD.com}
}
\email{hejie67@jd.com}

\author{Changping Peng}
\affiliation{%
  \country{JD.com}
}
\email{pengchangping@jd.com}

\author{Zhangang Lin}
\affiliation{%
  \country{JD.com}
}
\email{linzhangang@jd.com}

\author{Jingping Shao}
\affiliation{%
  \country{JD.com}
}
\email{shaojingping@jd.com}

\renewcommand{\shortauthors}{Yang Chen et al.}
\begin{abstract}
Online recommenders have attained growing interest and created great revenue for businesses. Given numerous users and items, incremental update becomes a mainstream paradigm for learning large-scale models in industrial scenarios, where only newly arrived data within a sliding window is fed into the model, meeting the strict requirements of quick response. However, this strategy would be prone to overfitting to newly arrived data. When there exists a significant drift of data distribution, the long-term information would be discarded, which harms the recommendation performance. Conventional methods address this issue through native model-based continual learning methods, without analyzing the data characteristics for online recommenders. To address the aforementioned issue, we propose an incremental update framework for online recommenders with Data-Driven Prior (DDP), which is composed of Feature Prior (FP) and Model Prior (MP). The FP performs the click estimation for each specific value to enhance the stability of the training process. The MP incorporates previous model output into the current update while strictly following the Bayes rules, resulting in a theoretically provable prior for the robust update. 
In this way, both the FP and MP are well integrated into the unified framework, which is model-agnostic and can accommodate various advanced interaction models. 
Extensive experiments on two publicly available datasets as well as an industrial dataset demonstrate the superior performance of the proposed framework.
\end{abstract}





\maketitle

\section{Introduction}
With the rapid development of Web applications, recommender systems have become ubiquitous to solve the serious information overload problem, with the aim of locating potentially preferred items among numerous items for users~\cite{chen2022cache,chen2022fast,chen2022learning,lian2020personalized}. Many e-commerce companies have generated significant revenue from recommender systems, where a higher click-through rate corresponds to a higher probability of greater benefit. Thus, CTR prediction plays a vital role in nowadays online recommendations.

In recent years, in order to better fit in the large-scale target data, expressive models are proposed to capture the complex interactions between multiple features~\cite{song2019autoint,liu2020autofis}, where the models tend to go more complex to fully capture the multiple high-order combinatorial features. However, these models require massive computational resources under the tremendous number of users and items, preventing quick online deployment and updates for industrial online recommendations. A mainstream framework is to feed models with the incremental data, i.e., newly arrived data within the sliding window, to continuously train the latest model rather than training from scratch~\cite{wang2020practical,cai2022reloop}. 
This strategy significantly reduces time overhead and adapts to dynamics in online data distribution.

However, when the distribution of online data changes significantly, this framework is prone to overfitting recent data. For example, considering the scenarios of large promotions, such as Double 11 shopping carnival and Black Friday, the specific commodities will receive a lot of exposure in a short period, resulting in a different distribution from previously collected feedback. The training framework would suffer from this and be inclined to these recent exposures. As a result, the model pays too much attention to the newly arrived data and gradually ignores the long-term information about user interests, limiting the recommendation performance. 
Moreover, due to the relatively decreased exposure ratio of long-tail items, the model pays less attention to these items, which aggravates the long-tail effect.

Existing studies apply continual learning to alleviate this issue for online recommenders. IncCTR~\cite{wang2020practical} adopts distillation logit to guide the latest model while ASMG~\cite{peng2021learning} incorporates the meta-learners to fit in recent models. 
To summarize, these studies tackle the issue by utilizing model-based prior to directly predicting CTR for each instance. However, these studies adopt strategies inspired by conventional continual learning, without analyzing the data characteristic in online recommendations. 
The extreme data sparsity and feature diversity are two prominent characteristics. 
Under the numerous numbers of users and items, user clicks are extremely sparse, which makes it difficult to accurately estimate user preference. 
Recent research~\cite{wu2021adversarial} suggests that each item requires around 10,000 impressions for convergence. 
Limited impressions hinder robust estimation during incremental updates.
The recent success of CTR prediction can be attributed to the use of complex features, where tail items with similar features to popular items obtain more accurate estimates. 
Additionally, considering that features are also the most important factor affecting the model effect~\cite{nadler2005prediction,wang2022autofield}, we are motivated to explicitly incorporate the feature prior to enhancing the performance.

To this end, we propose a robust and unified incremental learning framework with data-driven prior(DDP) to improve the performance under the nowadays mainstream training framework. It incorporates the feature prior in an end-to-end manner and provides a more theoretically provable model prior. More concretely, the Feature Prior is intended to explicitly estimate the average CTR of the specific feature value. The distribution of CTR value at the feature granularity presents more stable than the distribution at the instance level, since the data are more aggregated for the features. FP finnaly acts as auxiliary feature information and provides a more stable learning for model updates, thus benefitting the optimization for long-tail items. 
Furthermore, depending on the Bayes rules, we develop the Model Prior, which approximates the posterior on the complete data by maximizing the likelihood function on the incremental data and lowering the distance to the prior model.
Accordingly, the output from previous models can be easily integrated into the framework to achieve the model prior, where the output of the previous model is used to supervise the current model.

Our main contribution can be summarized as follows:
\begin{itemize}[leftmargin=*]
    \item We propose a robust and unified incremental update framework with data-driven prior to alleviate the overfitting problem on recent data, which is model-agnostic and can be optimized in an end-to-end manner.
    \item FP is designed to estimate the average CTR value for features, benefitting from a more stable distribution at the feature-level. This module can help CTR model more stable training and more accurate CTR estimation on long-tail items.
    \item MP approximates the posterior on the complete data by optimizing the likelihood function on the incremental data and reducing the distance to the previous model.
    \item Exhaustive experiments are conducted on two public dataset to demonstrate the superiority of the proposed framework. Both offline and online experiments are conducted on an industrial dataset schow the effectiveness in real scenarios.
\end{itemize}

\section{Related Work} 

\subsection{CTR Prediction Models}
Click-Through Rate (CTR) prediction plays an important role for online recommenders, such as online advertising, and numerous studies have investigated it in recent decade~\cite{cheng2016wide,wang2017deep,guo2017deepfm,pan2018field,song2019autoint}. These methods mainly vary in the way of capturing the feature interactions. Wide\&Deep~\cite{cheng2016wide} firstly combines the shallow linear layers and deep neural networks to simultaneously capture feature expression at different levels of features. Motived by Wide\&Deep, DeepFM~\cite{guo2017deepfm} replaces the wide part with the factorization machine models to better capture the second-order interactions between features. DCN~\cite{wang2017deep} explicitly constructs high-level combinatorial relations of features through Cross Network. FwFM~\cite{pan2018field} effectively captures the heterogeneity of field on second-order interaction features. AutoInt~\cite{song2019autoint} automatically learn the high-order feature interactions of input features with a multi-head self-attention neural network to explicitly model feature interactions in a low-dimensional space. Another mainstream direction is to model the historical behavior of users to better figure out user interests. 
DIN~\cite{zhou2018deep} obtains the user's interest in the target item by modeling the interest distribution of a user's browsing sequence under a given target item. 
Based on DIN, DIEN~\cite{zhou2019deep} investigates how user interests fluctuate over time and improves user interest capture by simulating the evolution of interests over time in the user sequence.

\subsection{Incremental Update for Online Recommenders}
Limited by the computational time and resources, the models are trained with streaming data, resulting in the incremental update for online recommendation. This incremental training framework has been widely used for industrial scenarios and attracted growing interest in recent years~\cite{wang2020practical,cai2022reloop,peng2021learning,guan2022deployable,mi2020ader,xia2022fire}.


Here we briefly introduce a few papers that are most relevant to our work. IncCTR~\cite{wang2020practical} uses the previous model as a teacher and supervises the learning of the current model through the commonly-used distillation loss. Motivated by~\cite{wang2020practical}, ReLoop~\cite{cai2022reloop} uses the hinge loss to guide the current model to perform better than the previous model by treating the performance of the previous model as the margin value. ASMG~\cite{peng2021learning} draws the idea of meta-learning and generates a better serving model from a sequence of historical models via a meta generator, so as to overcome the problem of overfitting and forgetting.

\section{Methodology}

\subsection{Preliminaries}
Assume each instance $(\mathcal{X},y)$ records a piece of impression, where $\mathcal{X}$ denotes features with $M$ fields, and $y \in \{0,1\}$ indicates whether the user clicks the item. After one-hot or multi-hot encoding for the features, each instance can be converted into $\mathbf{x}=[\mathbf{x}_1, ..., \mathbf{x}_M]$ where $\mathbf{x}_j $ denotes the encoding vector for the $j$-th field. The CTR-prediction model then learns a mapping $\hat{y} = f_{\theta} (\mathbf{x})$ to estimate the probability of user click. 
Previous models can be summarized with three parts: embedding module, interaction module and loss function. The embedding module convert encoding into embedding vector $\mathbf{e}=[\mathbf{e}_1, ..., \mathbf{e}_M]$, where $\mathbf{e}_j \in \mathbb{R}^d$ is obtained through the embedding matrix $\bm{E} \in \mathbb{R}^{N \times d}$. $N$ denotes the number of sparse feature values. The interaction module attempts to capture the complex relationships between features and to predict the probability of the user click through the sigmoid function, i.e., $\hat{y}= f_{\theta}(\mathbf{x}) = \sigma\left(\text{Interaction\_Module}(\mathbf{e})\right)$. $\theta$ denotes the set of parameters and $f(\cdot)$ summarizes all calculations.
Finally, the model is optimized through the binary cross entropy loss $\mathcal{L(\mathcal{D}; \theta)}$.

Considering the industrial scenarios under large scales, 
the incremental update becomes a popular framework, where the model is updated only with newly arrived data $\mathcal{D}_t$. $\mathcal{D}_t$ denotes the data collected from the period $t$, whereby a period can be a certain length of time (e.g., an hour, a day). $\mathcal{H}_{t}=\mathcal{D}_1 \cup \mathcal{D}_2 ,..., \cup \mathcal{D}_{t}$ denotes the historical training data up to the period $t$. Correspondingly, $\theta_{t-1}$ denotes model parameters updated in time $t-1$. The incremental update receives the newly arrived data $\mathcal{D}_{t}$ and previous model $\theta_{t-1}$, updating the model parameters, i.e., $\min \mathcal{L}(\mathcal{D}_{t}, \theta_{t-1}; \theta)$. It is obvious that the models are prone to overfitting to recent data when there is a drift of online data, and long-term user interest would be discarded under the loop of the incremental update. 

\begin{figure}[t]
    \centering
    
    \includegraphics[width=0.95\columnwidth]{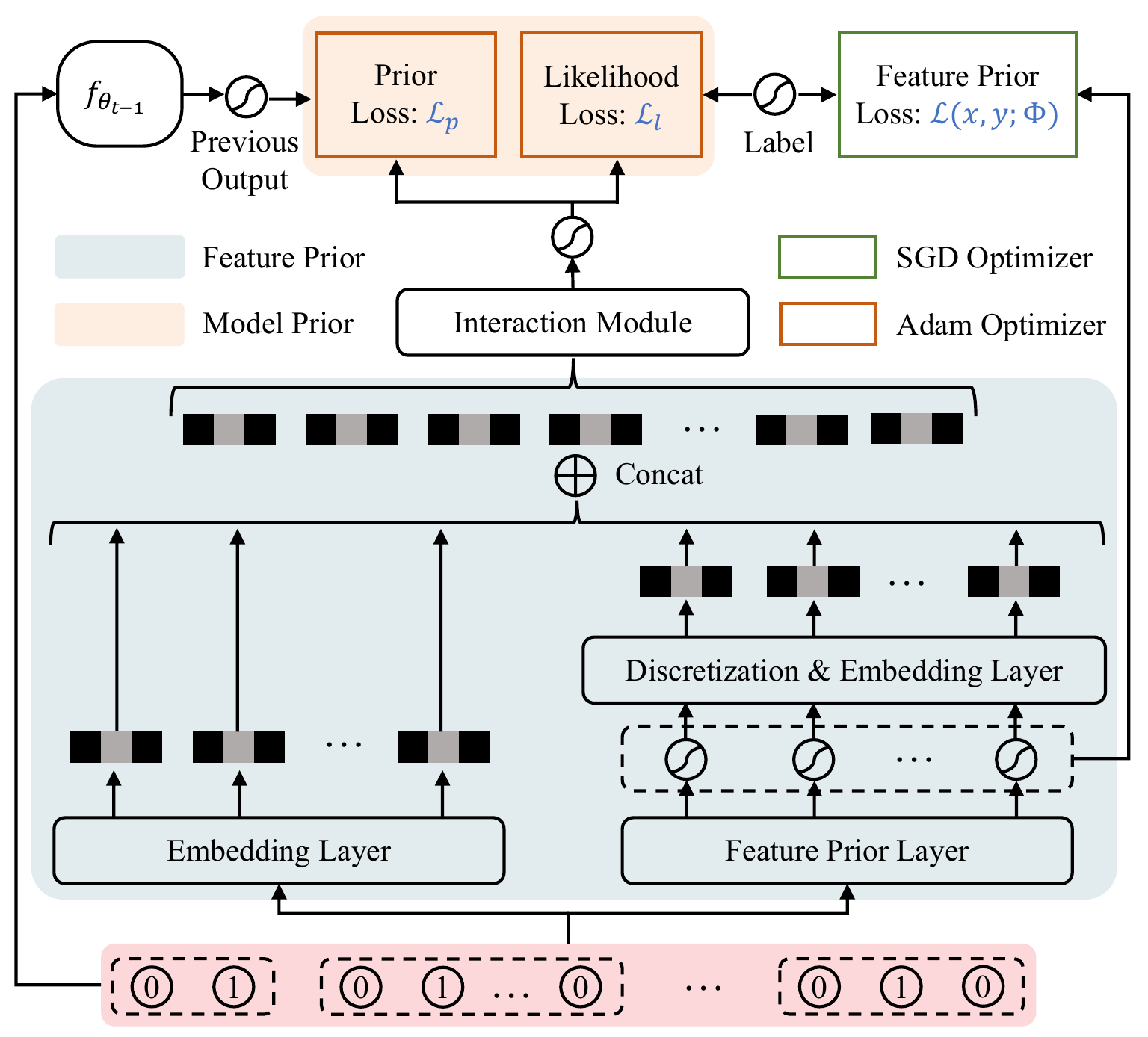} 
    \vspace{-0.3cm}
    \caption{The Architecture of DDP.}
    \label{fig::method}
    \vspace{-0.7cm}
\end{figure}

\subsection{Framework Overview}
We propose an unified framework, namely incremental update framework for online recommenders with Data-Driven Prior (DDP), which consist of two important components, as shown in Figure~\ref{fig::method}:

\noindent (1) \textbf{Feature Prior (FP)} estimates the average CTR of the specific value of each feature field, motivated by a more stable distribution at the feature level rather than the frequent dynamics at the instance level. The objective of FP is to help model learn more robust and have more accurate CTR estimates with long-tail data.

\noindent (2) \textbf{Model Prior (MP)} provides a more robust update depending on the Bayes rules, which acts as a regularizer to minimize the distance between the incremental update and the training with whole data. The purpose of MP is to assist the model in learning stably and avoiding over-fitting on incremental data.

These two data-driven parts can be easily integrated into existing advanced models, resulting in a model-agnostic and general framework. Furthermore, the framework can be updated in an end-to-end manner, showing its feasibility for online recommenders.

\begin{figure}
\centering
    \includegraphics[width=0.8\columnwidth]{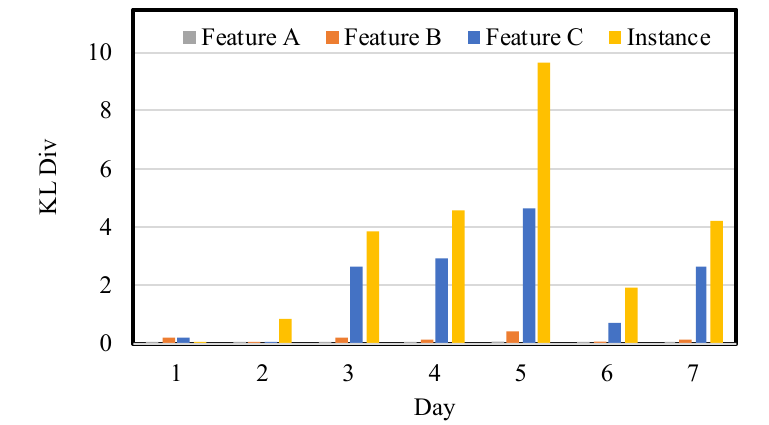} 
    \vspace{-0.4cm}
    \caption{KL-Distance w.r.t. statistical CTR of seven consecutive days. The KL-Distance measures the distance between the CTR value within the day and average value.}
    \label{fig::kl_div}
    \vspace{-0.3cm}
\end{figure}

\subsection{Feature Prior (FP)}
Previous studies adopt the native continual learning methods, by implementing the model-based information to predict the CTR for each instance. A major concern is the extreme data sparsity aggravated by incremental update. Model parameters are considerably sensitive to these data, where sparse data raises the instability of models, resulting in the overfitting to recent data. 

Intuitively, the feature data among the instances are more frequent, which acts more stable than instances. To further demonstrate it, we summarize the change of CTR values within seven consecutive days at the feature-level and instance-level in Figure~\ref{fig::kl_div}. Specifically, we select three features and corresponding instances. Feature C is sparser than Feature A and Feature B. The KL-distance is calculated with $KL=\sum_{i=0}^{1} q(i) \log \frac{q(i)}{p(i)}$, where $
q(i)$ and $p(i)$ denote the statistical CTR of features (instances) within the day and average value respectively. 
As shown in Figure~\ref{fig::kl_div}, the change of CTR with respect to the instances behaves more frequently than that of features, which brings difficulties for the update. Thus, we are motivated to design a module to estimate the CTR values for the features, which is fed into CTR model as stable and useful information, with the aim of improving the performance of recommenders. This feature-level value can generalize to tailed items, allowing CTR models to estimate long-tail characteristics more accurately. To this end, we propose the Feature Prior (FP), which can maintain long-term prior information for each feature, enabling a more stable expression of long-tail characteristics.

As aforementioned, each instance is represented with the one-hot encoding or multi-hot encoding:
$\mathbf{x} = [\mathbf{x}_1, \mathbf{x}_2, ..., \mathbf{x}_M]$
where $M$ denotes feature fields and $\mathbf{x}_i$ denotes the feature encoding of the field $i$. Notice all features are transformed into one-hot vectors here for simplicity. Apart from the dense high-dimensional embedding matrix, we design a feature prior layer to estimate the average CTR of each feature value as follows:
\begin{displaymath}
    \hat{s}_i = \sigma (\mathbf{C}^\top \mathbf{x}_i)
\end{displaymath}
where $\mathbf{C} \in \mathbb{R}^{N \times 1}$ denotes the feature prior embedding vector of all sparse features, $N$ denotes the number of sparse feature values and $\sigma$ is the sigmoid function. A binary cross entropy loss is applied to update the embedding vectors by maximum likelihood estimation: 
\begin{equation}
\footnotesize
    \label{eq::loss_f}
    \mathcal{L}(\mathbf{x}, y; \Phi) 
    = -\mathbb{E}_{{\mathbf{x},y} \sim \mathcal{D}_{t}} \left[ \frac{1}{M} \sum_i^M y \log(\hat{s}_i) + (1-y) \log(1-\hat{s}_i) \right]
\end{equation}
where $y$ denotes the label of the instance, $\mathcal{D}_t$ denotes incremental data and $\Phi$ denotes the parameters of the feature prior layer. Based on Eq~\eqref{eq::loss_f}, Feature Prior Loss, we can estimate the CTR values for each feature, which further aids the robust training.

Thus, we obtain a continuous estimated CTR value for each specific value. In order to integrate with the general embedding layers, i.e, the original feature values, the discretization strategy is adopted here, i.e., $\mathbf{s}_i = \mathrm{One\_Hot}(\lfloor B \cdot \sqrt{\hat{s_i}} \rfloor)$, where $B$ denotes the number of bins with equal width for discretization. Since the statistics CTR value is closer to 0 in real scenarios, the square root function is used to lead the model to focus more on smaller values.

After discretizing the estimated values, another embedding matrix $\bm{U}_i \in \mathbb{R}^{B \times d}$ is used to get the encoding vector for the $i$-th feature field. Specifically, $\bm{o}_i = \bm{U}_i^\top \mathbf{s}_i $. Thus, the low-dimensional dense embedding for the feature prior follows as:
\begin{displaymath}
    \mathbf{o} = [\mathbf{o}_1, \mathbf{o}_2, ..., \mathbf{o}_M]
\end{displaymath}
By concatenating vectors $\mathbf{e}$ and $\mathbf{o}$, we obtain the final embedding representation of features and their prior:
\begin{displaymath}
    \mathbf{e}' = \mathrm{Concat}([\mathbf{e}, \mathbf{o}]) = [\mathbf{e}_1, ..., \mathbf{e}_M, \mathbf{o}_1, ..., \mathbf{o}_M]
\end{displaymath}
where $\mathrm{Concat}(\cdot)$ denotes concat function. In this way, the feature prior is well integrated with the original embedding module, followed by any interaction module to capture the interactions between different features.

\subsection{Model Prior (MP)}
Empirically, the main reason for the model parameters overfitting to the incremental data distribution is that the model parameter optimization only considers the likelihood of the model parameter distribution on the incremental data, while neglecting the influence of previous data. Therefore, we conduct an in-depth analysis of the model optimization process, and finally simplify it to make our parameters fully learn the complete data distribution $\mathcal{H}_t = \{ \mathcal{H}_{t-1}, \mathcal{D}_{t}\}$ by fitting the likelihood on $ \mathcal{D}_{t}$ and constraining the distance between the output of current model and the output of previous model $\theta_{t-1}$, which is optimized by posterior with $\mathcal{H}_{t-1}$.

From a probabilistic perspective, our optimization goal is to optimize posterior parameter distribution $\theta_t$ given the complete data $\mathcal{H}_t$:
\begin{align}
    \notag
    \theta_{t} & = \mathop{\mathrm{argmax}}_{\theta} \log p(\theta | \mathcal{H}_t) \\
    \notag
    & = \mathop{\mathrm{argmax}}_{\theta} \left( \log p(\mathcal{H}_t | \theta) + \log p(\theta) - \log p(\mathcal{H}_t) \right) \\
    \notag
    & = \mathop{\mathrm{argmax}}_{\theta} \Big( \log p(\mathcal{D}_t | \theta) + \log p(\mathcal{H}_{t-1} | \theta) \\
    \notag
    & \qquad \qquad \qquad + \log p(\theta) - \log p(\mathcal{H}_{t-1}) - \log p(\mathcal{D}_{t}) \Big) \\
    \notag
    & = \mathop{\mathrm{argmax}}_{\theta}  \left( \log p(\mathcal{D}_{t} | \theta) + \log p(\theta | \mathcal{H}_{t-1}) - \log p(\mathcal{D}_{t}) \right) \\
    \label{eq::loss_two_part}
    &= \mathop{\mathrm{argmax}}_{\theta}  \left( \log p(\mathcal{D}_{t} | \theta) + \log p(\theta | \mathcal{H}_{t-1}) \right)
\end{align}
According to Eq~\eqref{eq::loss_two_part}, the objective is to maximize the likelihood $\log p(\mathcal{D}_{t} | \theta) $ and the posterior estimation $\log p(\theta | \mathcal{H}_{t-1})$ given the historical training data. Next we introduce the solutions respectively.
Regarding the likelihood $p(\mathcal{D}_{t} | \theta)$, by assuming that the user click follows a binomial distribution with the probability of $f_\theta(\mathbf{x})$, the objective function deduces to the binary cross-entropy loss: 
\begin{equation}
    \mathcal{L}_{l} = - \mathbb{E}_{{\mathbf{x},y} \sim \mathcal{D}_{t}} \left[ y \log(f_\theta(\mathbf{x})) + (1-y) \log(1-f_\theta(\mathbf{x})) \right]
    \label{eq::likellihood_loss}
\end{equation}
where $y\in \{0,1 \}$ denotes the click label and $f$ summarizes the model output. $f_\theta$ denotes the output parameterized by $\theta$. This objective function guides the model to fully fit the incremental data $\mathcal{D}_t$.

Regarding the posterior $p(\theta | \mathcal{H}_{t-1})$, it is intractable to directly obtain the solution. EWC~\cite{kirkpatrick2017overcoming} adopts the Laplace approximation to regularize the distance between parameter space $\theta$ and the optimal parameter space $\theta_{t-1}$ obtained given the historical training data $\mathcal{H}_{t-1}$, which acts as a surrogate loss to optimize $\log p(\theta | \mathcal{H}_{t-1})$. MRNFS~\cite{benjamin2018measuring} further simplifies the parameter space into the output of the models, with the aim of more stable learning to alleviate the strong constraints at the parameter level, achieving better performances. Thus, we follow these studies and get the following 
objective function:
\begin{align}
    \mathcal{L}_{p} &= \mathbb{E}_{\mathbf{x} \sim \mathcal{D}_{t}} \vert f_{\theta}(\mathbf{x}) - f_{\theta_{t-1}}(\mathbf{x}) \vert^2 
    \label{eq::prior_loss}
\end{align}
where $f_\theta$ stands for current learning model and $f_{\theta_{t-1}}$ denotes the previous well-learned model given $\mathcal{H}_{t-1}$.

Finally, the objective function follows: 
\begin{equation}
    \label{eq::final_loss}
        \mathcal{L}(\mathbf{x}, \mathbf{s}, y, \theta_{t-1}; \Theta) = \mathcal{L}_{l} + \frac{\lambda}{2} \mathcal{L}_{p}
\end{equation}
where $\lambda$ is the coefficient to regulate the relationship between the two losses, $\mathbf{s} = [\mathbf{s}_1, \mathbf{s}_2, ..., \mathbf{s}_M]$ denotes the output of feature prior layer and $\Theta$ denotes the other parameters except feature prior layer. By integrating with Eq~\eqref{eq::loss_f} of the Feature Prior (FP), we can update the whole framework in an end-to-end manner, without changing the structure of the important interaction modules. The whole process is detailed in  Alg~\ref{alg::train_model}. Note that if we optimize Eq~\eqref{eq::loss_f} with the Adam optimizer~\cite{kingma2015adam}, the feature prior layer will overfit within a short period due to the gradient vanish~\cite{wilson2017marginal}, which can not capture the change of data distribution continuously. Thus, we use the SGD optimizer to update the feature prior layer while using Adam optimizer for other parts.

\begin{algorithm}[t]
    \LinesNumbered
	\KwIn{Incremental Data $\mathcal{D}_{t}=\{\mathbf{x}, y\}$, The latest model parameters $\theta_{t-1}$}
    \KwOut{$\theta = \{\Phi, \Theta\}$} 
    \tcp{\footnotesize{$\Phi$ is the parameter of feature prior layer and $\Theta$ is the other parameters}}
	$\theta = \theta_{t-1}$; \\
	\For{$\{\mathbf{x}, y\}$ \rm{in} $\mathcal{D}_{t}$}{
	   Get $\mathbf{s} = [\mathbf{s}_1, \mathbf{s}_2, ..., \mathbf{s}_M]$ according to the Feature Prior Layer\;
	   Obtain $\mathcal{L}(\mathbf{x}, y; \Phi)$ through Eq~\eqref{eq::loss_f}\;
	   Obtain $\mathcal{L}(\mathbf{x}, \mathbf{s}, y, \theta_{t-1}; \Theta)$ through Eq~\eqref{eq::final_loss}\;
	   Optimize $\Phi$ using $\mathcal{L}(\mathbf{x}, y; \Phi)$ with SGD\;
	   Optimize $\Theta$ using $\mathcal{L}(\mathbf{x}, \mathbf{s}, y, \theta_{t-1}; \Theta)$ with Adam\;
	}
	\caption{DDP}
	\label{alg::train_model}
\end{algorithm}

\section{Experiments on Public Datasets}
\subsection{Experimental Settings}
\subsubsection{Datasets}
Two datasets, Criteo and CIKM2019, are adopted here for comparisons, whose statistics are summarized in Tabel~\ref{tab::data}.
\begin{itemize}[leftmargin=*]
    \item Criteo\footnote{\url{https://www.kaggle.com/c/criteo-display-ad-challenge}}, which is widely used for CTR prediction, is a collection of user traffic logs from Criteo for seven consecutive days. 
    \item CIKM2019\footnote{\url{https://tianchi.aliyun.com/competition/entrance/231721/introduction?lang=en-us}} is a public dataset from an e-commerce company. We follow the previous work~\cite{zhu2021learning} to transform the original label into binary values, where 1/0 indicates whether a user has bought an item or not. It consists of behavior logs for 16 consecutive days.
\end{itemize}
    To simulate the online settings of online industrial scenarios, the public data are fed into models in a streaming way. More concretely, within the logs of $T$ days, the proceeding data of $w$ days is used to warm up the model where the model is incremental updated for the next consecutive days with the time period as a day. Finally, the performance is tested on the last day after $T-w-1$ trails.
    For Criteo, $T=7$ and $w=3$. For CIKM2019 dataset, $T=16$ and $w=8$.
    
\begin{table}
    \centering
    \caption{Statistics of datasets. M represents a million.}
    \vspace{-0.3cm}
    \begin{tabular}{c|r|r|r}
        \toprule
        Dataset & \#Instances & \#Field & Time Duration \\
        \midrule
        Criteo & 45.84M & 39 & 7 days \\
        CIKM2019 & 58.75M & 8 & 16 days \\
    \bottomrule
    \end{tabular}
    \label{tab::data}
\end{table}

\begin{table}[t]
	\centering
    \caption{Overall performance with Baselines on the Criteo and CIKM2019 datasets. FP represents the feature prior. Notice DeepFM is chosen as the interaction module in DDP. Boldface denotes the highest score and underline indicates the best result of the baselines. $\star$ represents significance level $p$-value < 0.05 of comparing with the best baselines.}
    \vspace{-0.3cm}
    \scalebox{0.78}{
	\begin{tabular}{c|c|c|c|c|c|c}
		\toprule
		Dataset & \multicolumn{2}{c|}{Criteo} & \multicolumn{4}{c}{CIKM2019}  \\
        \midrule
        Data Split & \multicolumn{2}{c|}{All} & \multicolumn{2}{c|}{All} & \multicolumn{2}{c}{Long Tail} \\
		Metric & AUC $\uparrow$ & LogLoss $\downarrow$ & AUC $\uparrow$ & LogLoss $\downarrow$ & AUC $\uparrow$ & LogLoss $\downarrow$ \\ 
	    \midrule
		DNN & 0.7985 & 0.4546 & 0.7588 & 0.1234 & 0.7248 & 0.1212 \\
		DCN & 0.7982 & 0.4549 & 0.7588 & 0.1234 & 0.7250 & 0.1212 \\
		W\&D & 0.7985 & 0.4546 & 0.7587 & 0.1234 & 0.7249 & 0.1212 \\
		DeepFM & 0.7996 & 0.4537 & 0.7589 & 0.1234 & 0.7248 & 0.1212 \\
		FwFM & 0.7985 & 0.4546 & 0.7589 & 0.1234 & 0.7249 & 0.1212 \\
		AutoInt & 0.7986 & 0.4546 & 0.7587 & 0.1234 & 0.7247 & 0.1212 \\
		\midrule
		MECP & 0.8004 & 0.4531 & 0.7594 & 0.1235 & \underline{0.7278} & \underline{0.1209} \\
		IncCTR & 0.8007 & 0.4528 & 0.7597 & 0.1236 & 0.7275 & 0.1210 \\
		RLP & 0.8005 & 0.4531 & 0.7595 & \underline{0.1234} & 0.7273 & \underline{0.1209} \\
		ASMG & \underline{0.8020} & \underline{0.4517} & \underline{0.7599} & \underline{0.1234} & 0.7231 & 0.1213 \\
		\midrule
		FP & 0.8031 & 0.4507 & 0.7602 & \textbf{0.1232}$^\star$ & \textbf{0.7325}$^\star$ & \textbf{0.1204}$^\star$ \\
		DDP & \textbf{0.8038}$^\star$ & \textbf{0.4500}$^\star$ & \textbf{0.7608}$^\star$ & \textbf{0.1232}$^\star$ & 0.7317 & 0.1205 \\
	\midrule
        \% Improv. & \textbf{0.22\%} & \textbf{0.38\%} & \textbf{0.12\%} & \textbf{0.16\%} & \textbf{0.65\%} & \textbf{0.41\%} \\
        \bottomrule
	\end{tabular}}
	\label{table::general_res}
 \vspace{-0.5cm}
\end{table}

\subsubsection{Baselines} 
We choose the following competing methods as baselines for comparison: \textbf{IncCTR}~\cite{wang2020practical}, \textbf{RLP}~\cite{cai2022reloop}, \textbf{ASMG}~\cite{peng2021learning} and \textbf{MECP}~\cite{ling2017model}.
Moreover, since the proposed DDP is a model-agnostic framework, we apply DDP to six different deep CTR models: \textbf{DNN}, \textbf{DCN}~\cite{wang2017deep}, \textbf{W\&D}~\cite{cheng2016wide}, \textbf{DeepFM}~\cite{guo2017deepfm}, \textbf{FwFM}~\cite{pan2018field} and \textbf{AutoInt}~\cite{song2019autoint} to verify the compatibility of DDP with various interaction module. 

\subsubsection{Metric and Implementation Details}
We use the commonly used metrics in CTR prediction, \textbf{AUC} and \textbf{LogLoss}, for evaluation.
All the experiments are repeated 5 times by changing
the random seeds.
The models are trained within mini batch, where the learning rate is tuned over $\{1e-6, 5e-5, ..., 1e-2 \}$. The weight of $L_2$ regularization is tuned over $\{1e-6, 5e-5, ..., 1e-3 \}$. The embedding size is set to 16 and the hidden layers for the deep networks is fixed to $[200,200,200]$. All of these models select the parameters with best performances in terms of AUC.

\subsection{Overall Performance}
We compare the performance of different deep CTR models for incremental update and various embedding-based deep CTR models. FP represents that only Feature Prior is included under the DeepFM interaction module. The results of our proposed approaches and baselines on two public datasets are shown in Table \ref{table::general_res}, from which we have the following observations:

\begin{itemize}[leftmargin=*]
    \item Our methods: FP and DDP consistently outperform other different deep CTR methods for incremental update and various embedding-based deep CTR models.
    Compared with the best basline ASMG, our method DDP achieves a relative improvement of 0.22\% and 0.12\% with respect to AUC on Criteo and CIKM2019 respectively. Notice, even a slight 0.1\% improvement on the CTR prediction task is significant~\cite{guo2017deepfm,wang2017deep}.
    \item Introducing the feature prior raises more improvements on the long-tail items than the short-hot items. We split the items into short-hot and long-tail according to the occurrence frequency of items. The 20\% items with higher frequency are defined as short hot, and the remaining 80\% are defined as long tail. The final ratio of short-hot and long-tail items of instances is about 8:2. In general, compared with DeepFM, our DDP improves AUC 0.95\% on long-tail items and 0.11\% on short hot items respectively. The reason lies in that the 
   FP performs more stable and obtains more general prior information, which is more helpful to provide a more accurate estimation on tailed items with limited training instances. This finding is consistent with the previous conclusion that the use of prior information is a very effective method to solve the long tail problem~\cite{cao2020open,cao2021learning,wang2020one}.
\end{itemize}

\begin{figure}[t]
    \centering
    \subfigure[AUC]{
    \includegraphics[width=0.48\columnwidth]{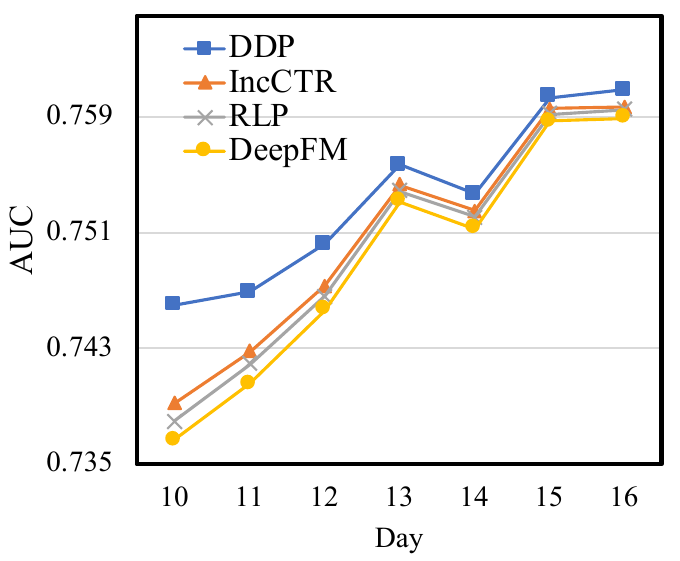}}
    \hspace{-0.3cm}
    \subfigure[LogLoss]{
    \includegraphics[width=0.48\columnwidth]{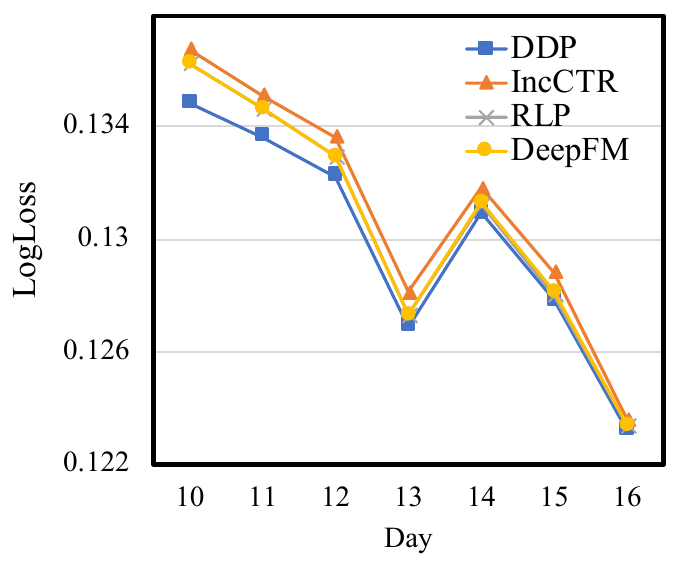}}
    \vspace{-0.35cm}
    \caption{Performances of Consecutive Days under Incremental Update on CIKM2019 Dataset}
    \label{fig:performance_consecutive_days}
    \vspace{-0.45cm}
\end{figure}

\begin{figure}[t]
    \centering
    \subfigure[AUC on Criteo]
        {
        \includegraphics[width=0.49\columnwidth]{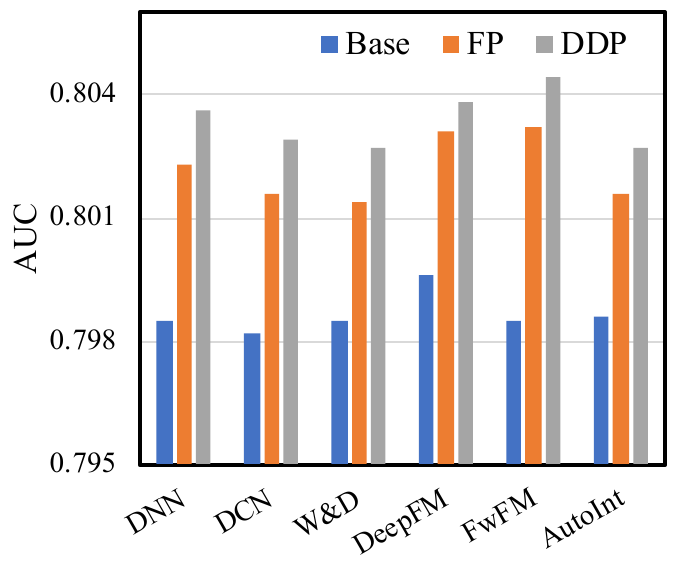}
        }
    \hspace{-0.4cm}
    \subfigure[LogLoss on Criteo]{
        \includegraphics[width=0.49\columnwidth]{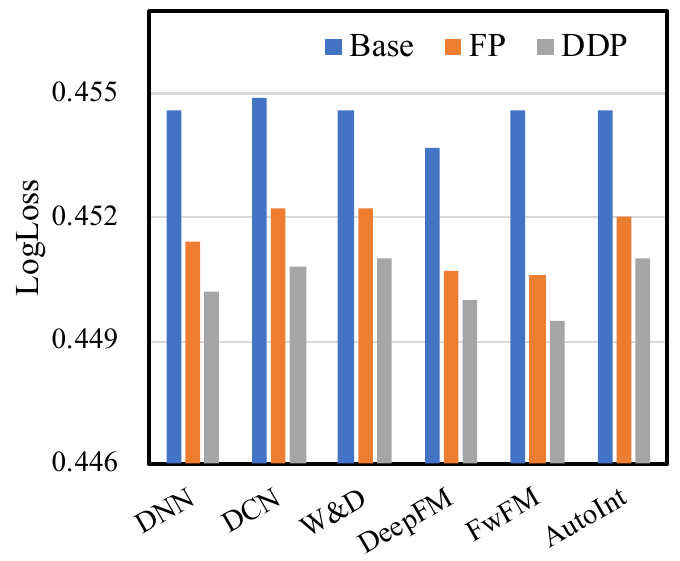}
        }
    \vspace{-0.35cm}
    \caption{Varying Interaction Modules}
    \label{fig:vary_interaction_model}
    \vspace{-0.35cm}
\end{figure}

\subsection{Performances on Consecutive Increments}
Considering that the real scenarios, where models are updated with the newly arrived data in loops, we mimic the online setting with consecutive days to explore the performance during the training loops. Experiments are carried out on the CIKM2019 dataset, where we train a warmed-up model using data from the preceding 8 days. The model is incrementally updated till the data before the test day, then tested on test day. For example, in the tenth day, the data before the tenth day is fed into the model incrementally and the test data is the tenth day.

As shown in Figure~\ref{fig:performance_consecutive_days}, DDP consistently outperforms the competing baseline IncCTR and RLP on the all seven tested days in terms of both AUC and LogLoss. This indicates the superiority of adopting the data-driven prior, which achieves a more robust update for recommenders. In the early stages, there is a large gap between our method and other baselines, which is primarily due to the fact that our FP can assist to better express the feature when the quantity of each feature is limited.

\subsection{Varying Interaction Modules}
Since the DDP is a model-agnostic framework and is easily plugged into various deep CTR models, we conduct experiments with six different interaction modules to verify the compatibility of the proposed method. The experimental results are shown in Figure~\ref{fig:vary_interaction_model}.

FP consistently improves the performance on all interaction modules, achieving an average 0.44\% improvement on AUC and 0.66\% on LogLoss. DDP further improves the performance with the additional MP. The improvements demonstrate the capability of the proposed method being suitable to any modern interaction modules and the superiority of alleviating the overfitting problem in incremental update for online recommenders.

\subsection{Deep Analysis of Feature Prior (FP)}
\begin{figure}[t]
    \centering
    \includegraphics[width=0.75\columnwidth]{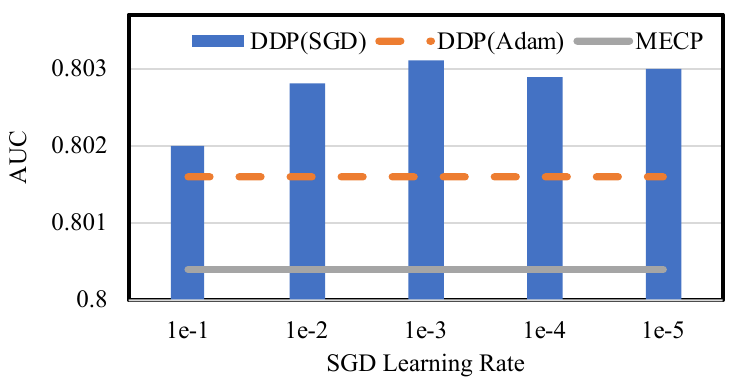}
    \vspace{-0.3cm}
    \caption{FP Optimized with SGD }
    \label{fig:diff_optimizer}
    \vspace{-0.35cm}
\end{figure}

As we claim aforementioned, the FP is optimized with SGD optimizer while MP is updated with Adam optimizer. We conduct experiments to compare the performance of updating FP with SGD and Adam optimizers, as shown in Figure~\ref{fig:diff_optimizer}. The performance under SGD is tuned over different learning rate. The orange dashed line represents the best performance with Adam optimizer. The grey line represents the performance of the competing method MECP, which adopts the previously trained CTR of features. 

The results demonstrate the superiority of integrating the FP in the end-to-end framework and we have the following observation: 1) DDP, capturing the dynamics of features along with instances, shows better performances than MECP, indicating the importance of constantly capturing data distribution at the feature level. 2) SGD outperforms Adam over a range of learning rates. 
This suggests that the decreasing learning rate of Adam throughout training hinders the continuous learning of the FP module.
3) Lower SGD learning rates lead to better performance by preserving long-term knowledge and ensuring parameter stability for feature CTR values.

\section{Experiments on Industrial Dataset}
The industrial dataset is a collection of logs from the homepage streaming data, which is from an online advertising platform of a famous e-commerce company. 
It consists of 7 billion instances over 121 consecutive days. 
We use the 120-day data for incremental training directly. 
The performance is evaluated on the data of the last day, as reported in Table~\ref{table::industry_offline}.
DDP outperforms the base models, achieving a relative 0.58\% improvement of AUC, even though the model has been already optimized with 7 billion logs. This supports our point that under the environment of rapid changes in data distribution, how maintaining the relatively stable prior information of the model and feature is of great significance to improve the overall performance of the model. 

Furthermore, we conduct online A/B test to validate the effectiveness of the proposed framework DDP, which is deployed on a mainstream advertising position on the homepage. There are around ten million impressions each day and we allocate 5\% of all traffic to DDP for 7 days. The results is reported in Table~\ref{table::industry_offline}. DDP increases CTR at an average 1.99\% improvement and increases eCPM (effective cost per mille) at 2.97\%, indicating significant improvements for online recommenders. 


\begin{table}
	\centering
	   \caption{Experiments on the Industrial Dataset}
        \vspace{-0.4cm}
		\begin{tabular}{c|c|c}
			\toprule
			Offline & AUC $\uparrow$ & LogLoss $\downarrow$   \\ 
			\midrule
			Base Model & 0.7382 & 0.1208 \\
			DDP & 0.7425 (+0.58\%) & 0.1199 (-0.75\%)  \\
			\bottomrule
            \toprule
            Online & CTR Gain & eCPM Gain \\
            \midrule
            DDP & +1.99\% & +2.97\% \\
            \bottomrule
		\end{tabular}
	\vspace{-0.4cm}
	\label{table::industry_offline}
\end{table}
 


\section{Conclusion}
In this paper, we propose an incremental update framework for online recommenders with Data-Driven Prior (DDP) to ease the overfitting problem on incremental data, which is model-agnostic and can be learned end-to-end. The proposed FP is intended to estimate the average CTR value for each feature, allowing for more stable training and more accurate CTR estimates for long-tail items. Further, the proposed MP approximates the posterior on the complete data by optimizing the likelihood function on the incremental data and reducing the distance with the previous model. Extensive experiments on two public datasets demonstrate the advantages of the proposed framework, while offline and online studies on an industrial dataset show its effectiveness.

\balance
\bibliographystyle{ACM-Reference-Format}
\bibliography{sample-base}


\begin{thebibliography}{30}


\ifx \showCODEN    \undefined \def \showCODEN     #1{\unskip}     \fi
\ifx \showDOI      \undefined \def \showDOI       #1{#1}\fi
\ifx \showISBNx    \undefined \def \showISBNx     #1{\unskip}     \fi
\ifx \showISBNxiii \undefined \def \showISBNxiii  #1{\unskip}     \fi
\ifx \showISSN     \undefined \def \showISSN      #1{\unskip}     \fi
\ifx \showLCCN     \undefined \def \showLCCN      #1{\unskip}     \fi
\ifx \shownote     \undefined \def \shownote      #1{#1}          \fi
\ifx \showarticletitle \undefined \def \showarticletitle #1{#1}   \fi
\ifx \showURL      \undefined \def \showURL       {\relax}        \fi
\providecommand\bibfield[2]{#2}
\providecommand\bibinfo[2]{#2}
\providecommand\natexlab[1]{#1}
\providecommand\showeprint[2][]{arXiv:#2}

\bibitem[Benjamin et~al\mbox{.}(2018)]%
        {benjamin2018measuring}
\bibfield{author}{\bibinfo{person}{Ari Benjamin}, \bibinfo{person}{David
  Rolnick}, {and} \bibinfo{person}{Konrad Kording}.}
  \bibinfo{year}{2018}\natexlab{}.
\newblock \showarticletitle{Measuring and regularizing networks in function
  space}. In \bibinfo{booktitle}{\emph{International Conference on Learning
  Representations}}.
\newblock


\bibitem[Cai et~al\mbox{.}(2022)]%
        {cai2022reloop}
\bibfield{author}{\bibinfo{person}{Guohao Cai}, \bibinfo{person}{Jieming Zhu},
  \bibinfo{person}{Quanyu Dai}, \bibinfo{person}{Zhenhua Dong},
  \bibinfo{person}{Xiuqiang He}, \bibinfo{person}{Ruiming Tang}, {and}
  \bibinfo{person}{Rui Zhang}.} \bibinfo{year}{2022}\natexlab{}.
\newblock \showarticletitle{ReLoop: A Self-Correction Continual Learning Loop
  for Recommender Systems}.
\newblock \bibinfo{journal}{\emph{Proceedings of the 45th International ACM
  SIGIR Conference on Research and Development in Information Retrieval}}
  (\bibinfo{year}{2022}).
\newblock


\bibitem[Cao et~al\mbox{.}(2020)]%
        {cao2020open}
\bibfield{author}{\bibinfo{person}{Ermei Cao}, \bibinfo{person}{Difeng Wang},
  \bibinfo{person}{Jiacheng Huang}, {and} \bibinfo{person}{Wei Hu}.}
  \bibinfo{year}{2020}\natexlab{}.
\newblock \showarticletitle{Open knowledge enrichment for long-tail entities}.
  In \bibinfo{booktitle}{\emph{Proceedings of The Web Conference 2020}}.
  \bibinfo{pages}{384--394}.
\newblock


\bibitem[Cao et~al\mbox{.}(2021)]%
        {cao2021learning}
\bibfield{author}{\bibinfo{person}{Yixin Cao}, \bibinfo{person}{Jun Kuang},
  \bibinfo{person}{Ming Gao}, \bibinfo{person}{Aoying Zhou},
  \bibinfo{person}{Yonggang Wen}, {and} \bibinfo{person}{Tat-Seng Chua}.}
  \bibinfo{year}{2021}\natexlab{}.
\newblock \showarticletitle{Learning relation prototype from unlabeled texts
  for long-tail relation extraction}.
\newblock \bibinfo{journal}{\emph{IEEE Transactions on Knowledge and Data
  Engineering}} (\bibinfo{year}{2021}).
\newblock


\bibitem[Chen et~al\mbox{.}(2022a)]%
        {chen2022fast}
\bibfield{author}{\bibinfo{person}{Jin Chen}, \bibinfo{person}{Defu Lian},
  \bibinfo{person}{Binbin Jin}, \bibinfo{person}{Xu Huang},
  \bibinfo{person}{Kai Zheng}, {and} \bibinfo{person}{Enhong Chen}.}
  \bibinfo{year}{2022}\natexlab{a}.
\newblock \showarticletitle{Fast variational autoencoder with inverted
  multi-index for collaborative filtering}. In
  \bibinfo{booktitle}{\emph{Proceedings of the ACM Web Conference 2022}}.
  \bibinfo{pages}{1944--1954}.
\newblock


\bibitem[Chen et~al\mbox{.}(2022b)]%
        {chen2022learning}
\bibfield{author}{\bibinfo{person}{Jin Chen}, \bibinfo{person}{Defu Lian},
  \bibinfo{person}{Binbin Jin}, \bibinfo{person}{Kai Zheng}, {and}
  \bibinfo{person}{Enhong Chen}.} \bibinfo{year}{2022}\natexlab{b}.
\newblock \showarticletitle{Learning Recommenders for Implicit Feedback with
  Importance Resampling}. In \bibinfo{booktitle}{\emph{Proceedings of the ACM
  Web Conference 2022}}. \bibinfo{pages}{1997--2005}.
\newblock


\bibitem[Chen et~al\mbox{.}(2022c)]%
        {chen2022cache}
\bibfield{author}{\bibinfo{person}{Jin Chen}, \bibinfo{person}{Defu Lian},
  \bibinfo{person}{Yucheng Li}, \bibinfo{person}{Baoyun Wang},
  \bibinfo{person}{Kai Zheng}, {and} \bibinfo{person}{Enhong Chen}.}
  \bibinfo{year}{2022}\natexlab{c}.
\newblock \showarticletitle{Cache-Augmented Inbatch Importance Resampling for
  Training Recommender Retriever}.
\newblock \bibinfo{journal}{\emph{Advances in Neural Information Processing
  Systems}}  \bibinfo{volume}{35} (\bibinfo{year}{2022}),
  \bibinfo{pages}{34817--34830}.
\newblock


\bibitem[Cheng et~al\mbox{.}(2016)]%
        {cheng2016wide}
\bibfield{author}{\bibinfo{person}{Heng-Tze Cheng}, \bibinfo{person}{Levent
  Koc}, \bibinfo{person}{Jeremiah Harmsen}, \bibinfo{person}{Tal Shaked},
  \bibinfo{person}{Tushar Chandra}, \bibinfo{person}{Hrishi Aradhye},
  \bibinfo{person}{Glen Anderson}, \bibinfo{person}{Greg Corrado},
  \bibinfo{person}{Wei Chai}, \bibinfo{person}{Mustafa Ispir}, {et~al\mbox{.}}}
  \bibinfo{year}{2016}\natexlab{}.
\newblock \showarticletitle{Wide \& deep learning for recommender systems}. In
  \bibinfo{booktitle}{\emph{Proceedings of the 1st workshop on deep learning
  for recommender systems}}. \bibinfo{pages}{7--10}.
\newblock


\bibitem[Guan et~al\mbox{.}(2022)]%
        {guan2022deployable}
\bibfield{author}{\bibinfo{person}{Renchu Guan}, \bibinfo{person}{Haoyu Pang},
  \bibinfo{person}{Fausto Giunchiglia}, \bibinfo{person}{Ximing Li},
  \bibinfo{person}{Xuefeng Yang}, {and} \bibinfo{person}{Xiaoyue Feng}.}
  \bibinfo{year}{2022}\natexlab{}.
\newblock \showarticletitle{Deployable and Continuable Meta-learning-Based
  Recommender System with Fast User-Incremental Updates}. In
  \bibinfo{booktitle}{\emph{Proceedings of the 45th International ACM SIGIR
  Conference on Research and Development in Information Retrieval}}.
  \bibinfo{pages}{1423--1433}.
\newblock


\bibitem[Guo et~al\mbox{.}(2017)]%
        {guo2017deepfm}
\bibfield{author}{\bibinfo{person}{Huifeng Guo}, \bibinfo{person}{Ruiming
  Tang}, \bibinfo{person}{Yunming Ye}, \bibinfo{person}{Zhenguo Li}, {and}
  \bibinfo{person}{Xiuqiang He}.} \bibinfo{year}{2017}\natexlab{}.
\newblock \showarticletitle{DeepFM: A Factorization-Machine based Neural
  Network for CTR Prediction}. In \bibinfo{booktitle}{\emph{IJCAI}}.
\newblock


\bibitem[Kingma and Ba(2015)]%
        {kingma2015adam}
\bibfield{author}{\bibinfo{person}{Diederik~P Kingma} {and}
  \bibinfo{person}{Jimmy Ba}.} \bibinfo{year}{2015}\natexlab{}.
\newblock \showarticletitle{Adam: A Method for Stochastic Optimization}. In
  \bibinfo{booktitle}{\emph{ICLR (Poster)}}.
\newblock


\bibitem[Kirkpatrick et~al\mbox{.}(2017)]%
        {kirkpatrick2017overcoming}
\bibfield{author}{\bibinfo{person}{James Kirkpatrick}, \bibinfo{person}{Razvan
  Pascanu}, \bibinfo{person}{Neil Rabinowitz}, \bibinfo{person}{Joel Veness},
  \bibinfo{person}{Guillaume Desjardins}, \bibinfo{person}{Andrei~A Rusu},
  \bibinfo{person}{Kieran Milan}, \bibinfo{person}{John Quan},
  \bibinfo{person}{Tiago Ramalho}, \bibinfo{person}{Agnieszka
  Grabska-Barwinska}, {et~al\mbox{.}}} \bibinfo{year}{2017}\natexlab{}.
\newblock \showarticletitle{Overcoming catastrophic forgetting in neural
  networks}.
\newblock \bibinfo{journal}{\emph{Proceedings of the national academy of
  sciences}} \bibinfo{volume}{114}, \bibinfo{number}{13}
  (\bibinfo{year}{2017}), \bibinfo{pages}{3521--3526}.
\newblock


\bibitem[Lian et~al\mbox{.}(2020)]%
        {lian2020personalized}
\bibfield{author}{\bibinfo{person}{Defu Lian}, \bibinfo{person}{Qi Liu}, {and}
  \bibinfo{person}{Enhong Chen}.} \bibinfo{year}{2020}\natexlab{}.
\newblock \showarticletitle{Personalized ranking with importance sampling}. In
  \bibinfo{booktitle}{\emph{Proceedings of The Web Conference 2020}}.
  \bibinfo{pages}{1093--1103}.
\newblock


\bibitem[Ling et~al\mbox{.}(2017)]%
        {ling2017model}
\bibfield{author}{\bibinfo{person}{Xiaoliang Ling}, \bibinfo{person}{Weiwei
  Deng}, \bibinfo{person}{Chen Gu}, \bibinfo{person}{Hucheng Zhou},
  \bibinfo{person}{Cui Li}, {and} \bibinfo{person}{Feng Sun}.}
  \bibinfo{year}{2017}\natexlab{}.
\newblock \showarticletitle{Model ensemble for click prediction in bing search
  ads}. In \bibinfo{booktitle}{\emph{Proceedings of the 26th international
  conference on world wide web companion}}. \bibinfo{pages}{689--698}.
\newblock


\bibitem[Liu et~al\mbox{.}(2020)]%
        {liu2020autofis}
\bibfield{author}{\bibinfo{person}{Bin Liu}, \bibinfo{person}{Chenxu Zhu},
  \bibinfo{person}{Guilin Li}, \bibinfo{person}{Weinan Zhang},
  \bibinfo{person}{Jincai Lai}, \bibinfo{person}{Ruiming Tang},
  \bibinfo{person}{Xiuqiang He}, \bibinfo{person}{Zhenguo Li}, {and}
  \bibinfo{person}{Yong Yu}.} \bibinfo{year}{2020}\natexlab{}.
\newblock \showarticletitle{Autofis: Automatic feature interaction selection in
  factorization models for click-through rate prediction}. In
  \bibinfo{booktitle}{\emph{Proceedings of the 26th ACM SIGKDD International
  Conference on Knowledge Discovery \& Data Mining}}.
  \bibinfo{pages}{2636--2645}.
\newblock


\bibitem[Mi et~al\mbox{.}(2020)]%
        {mi2020ader}
\bibfield{author}{\bibinfo{person}{Fei Mi}, \bibinfo{person}{Xiaoyu Lin}, {and}
  \bibinfo{person}{Boi Faltings}.} \bibinfo{year}{2020}\natexlab{}.
\newblock \showarticletitle{Ader: Adaptively distilled exemplar replay towards
  continual learning for session-based recommendation}. In
  \bibinfo{booktitle}{\emph{Fourteenth ACM Conference on Recommender Systems}}.
  \bibinfo{pages}{408--413}.
\newblock


\bibitem[Nadler and Coifman(2005)]%
        {nadler2005prediction}
\bibfield{author}{\bibinfo{person}{Boaz Nadler} {and} \bibinfo{person}{Ronald~R
  Coifman}.} \bibinfo{year}{2005}\natexlab{}.
\newblock \showarticletitle{The prediction error in CLS and PLS: the importance
  of feature selection prior to multivariate calibration}.
\newblock \bibinfo{journal}{\emph{Journal of Chemometrics: A Journal of the
  Chemometrics Society}} \bibinfo{volume}{19}, \bibinfo{number}{2}
  (\bibinfo{year}{2005}), \bibinfo{pages}{107--118}.
\newblock


\bibitem[Pan et~al\mbox{.}(2018)]%
        {pan2018field}
\bibfield{author}{\bibinfo{person}{Junwei Pan}, \bibinfo{person}{Jian Xu},
  \bibinfo{person}{Alfonso~Lobos Ruiz}, \bibinfo{person}{Wenliang Zhao},
  \bibinfo{person}{Shengjun Pan}, \bibinfo{person}{Yu Sun}, {and}
  \bibinfo{person}{Quan Lu}.} \bibinfo{year}{2018}\natexlab{}.
\newblock \showarticletitle{Field-weighted factorization machines for
  click-through rate prediction in display advertising}. In
  \bibinfo{booktitle}{\emph{Proceedings of the 2018 World Wide Web
  Conference}}. \bibinfo{pages}{1349--1357}.
\newblock


\bibitem[Peng et~al\mbox{.}(2021)]%
        {peng2021learning}
\bibfield{author}{\bibinfo{person}{Danni Peng}, \bibinfo{person}{Sinno~Jialin
  Pan}, \bibinfo{person}{Jie Zhang}, {and} \bibinfo{person}{Anxiang Zeng}.}
  \bibinfo{year}{2021}\natexlab{}.
\newblock \showarticletitle{Learning an Adaptive Meta Model-Generator for
  Incrementally Updating Recommender Systems}. In
  \bibinfo{booktitle}{\emph{Fifteenth ACM Conference on Recommender Systems}}.
  \bibinfo{pages}{411--421}.
\newblock


\bibitem[Song et~al\mbox{.}(2019)]%
        {song2019autoint}
\bibfield{author}{\bibinfo{person}{Weiping Song}, \bibinfo{person}{Chence Shi},
  \bibinfo{person}{Zhiping Xiao}, \bibinfo{person}{Zhijian Duan},
  \bibinfo{person}{Yewen Xu}, \bibinfo{person}{Ming Zhang}, {and}
  \bibinfo{person}{Jian Tang}.} \bibinfo{year}{2019}\natexlab{}.
\newblock \showarticletitle{Autoint: Automatic feature interaction learning via
  self-attentive neural networks}. In \bibinfo{booktitle}{\emph{Proceedings of
  the 28th ACM International Conference on Information and Knowledge
  Management}}. \bibinfo{pages}{1161--1170}.
\newblock


\bibitem[Wang et~al\mbox{.}(2017)]%
        {wang2017deep}
\bibfield{author}{\bibinfo{person}{Ruoxi Wang}, \bibinfo{person}{Bin Fu},
  \bibinfo{person}{Gang Fu}, {and} \bibinfo{person}{Mingliang Wang}.}
  \bibinfo{year}{2017}\natexlab{}.
\newblock \showarticletitle{Deep \& cross network for ad click predictions}.
\newblock In \bibinfo{booktitle}{\emph{Proceedings of the ADKDD'17}}.
  \bibinfo{pages}{1--7}.
\newblock


\bibitem[Wang et~al\mbox{.}(2020b)]%
        {wang2020one}
\bibfield{author}{\bibinfo{person}{Weitao Wang}, \bibinfo{person}{Meng Wang},
  \bibinfo{person}{Sen Wang}, \bibinfo{person}{Guodong Long},
  \bibinfo{person}{Lina Yao}, \bibinfo{person}{Guilin Qi}, {and}
  \bibinfo{person}{Yang Chen}.} \bibinfo{year}{2020}\natexlab{b}.
\newblock \showarticletitle{One-shot learning for long-tail visual relation
  detection}. In \bibinfo{booktitle}{\emph{Proceedings of the AAAI Conference
  on Artificial Intelligence}}, Vol.~\bibinfo{volume}{34}.
  \bibinfo{pages}{12225--12232}.
\newblock


\bibitem[Wang et~al\mbox{.}(2020a)]%
        {wang2020practical}
\bibfield{author}{\bibinfo{person}{Yichao Wang}, \bibinfo{person}{Huifeng Guo},
  \bibinfo{person}{Ruiming Tang}, \bibinfo{person}{Zhirong Liu}, {and}
  \bibinfo{person}{Xiuqiang He}.} \bibinfo{year}{2020}\natexlab{a}.
\newblock \showarticletitle{A practical incremental method to train deep ctr
  models}.
\newblock \bibinfo{journal}{\emph{arXiv preprint arXiv:2009.02147}}
  (\bibinfo{year}{2020}).
\newblock


\bibitem[Wang et~al\mbox{.}(2022)]%
        {wang2022autofield}
\bibfield{author}{\bibinfo{person}{Yejing Wang}, \bibinfo{person}{Xiangyu
  Zhao}, \bibinfo{person}{Tong Xu}, {and} \bibinfo{person}{Xian Wu}.}
  \bibinfo{year}{2022}\natexlab{}.
\newblock \showarticletitle{Autofield: Automating feature selection in deep
  recommender systems}. In \bibinfo{booktitle}{\emph{Proceedings of the ACM Web
  Conference 2022}}. \bibinfo{pages}{1977--1986}.
\newblock


\bibitem[Wilson et~al\mbox{.}(2017)]%
        {wilson2017marginal}
\bibfield{author}{\bibinfo{person}{Ashia~C Wilson}, \bibinfo{person}{Rebecca
  Roelofs}, \bibinfo{person}{Mitchell Stern}, \bibinfo{person}{Nati Srebro},
  {and} \bibinfo{person}{Benjamin Recht}.} \bibinfo{year}{2017}\natexlab{}.
\newblock \showarticletitle{The marginal value of adaptive gradient methods in
  machine learning}.
\newblock \bibinfo{journal}{\emph{Advances in neural information processing
  systems}}  \bibinfo{volume}{30} (\bibinfo{year}{2017}).
\newblock


\bibitem[Wu et~al\mbox{.}(2021)]%
        {wu2021adversarial}
\bibfield{author}{\bibinfo{person}{Kailun Wu}, \bibinfo{person}{Weijie Bian},
  \bibinfo{person}{Zhangming Chan}, \bibinfo{person}{Lejian Ren},
  \bibinfo{person}{Shiming Xiang}, \bibinfo{person}{Shuguang Han},
  \bibinfo{person}{Hongbo Deng}, {and} \bibinfo{person}{Bo Zheng}.}
  \bibinfo{year}{2021}\natexlab{}.
\newblock \showarticletitle{Adversarial Gradient Driven Exploration for Deep
  Click-Through Rate Prediction}.
\newblock \bibinfo{journal}{\emph{arXiv preprint arXiv:2112.11136}}
  (\bibinfo{year}{2021}).
\newblock


\bibitem[Xia et~al\mbox{.}(2022)]%
        {xia2022fire}
\bibfield{author}{\bibinfo{person}{Jiafeng Xia}, \bibinfo{person}{Dongsheng
  Li}, \bibinfo{person}{Hansu Gu}, \bibinfo{person}{Jiahao Liu},
  \bibinfo{person}{Tun Lu}, {and} \bibinfo{person}{Ning Gu}.}
  \bibinfo{year}{2022}\natexlab{}.
\newblock \showarticletitle{FIRE: Fast Incremental Recommendation with Graph
  Signal Processing}. In \bibinfo{booktitle}{\emph{Proceedings of the ACM Web
  Conference 2022}}. \bibinfo{pages}{2360--2369}.
\newblock


\bibitem[Zhou et~al\mbox{.}(2019)]%
        {zhou2019deep}
\bibfield{author}{\bibinfo{person}{Guorui Zhou}, \bibinfo{person}{Na Mou},
  \bibinfo{person}{Ying Fan}, \bibinfo{person}{Qi Pi}, \bibinfo{person}{Weijie
  Bian}, \bibinfo{person}{Chang Zhou}, \bibinfo{person}{Xiaoqiang Zhu}, {and}
  \bibinfo{person}{Kun Gai}.} \bibinfo{year}{2019}\natexlab{}.
\newblock \showarticletitle{Deep interest evolution network for click-through
  rate prediction}. In \bibinfo{booktitle}{\emph{Proceedings of the AAAI
  conference on artificial intelligence}}, Vol.~\bibinfo{volume}{33}.
  \bibinfo{pages}{5941--5948}.
\newblock


\bibitem[Zhou et~al\mbox{.}(2018)]%
        {zhou2018deep}
\bibfield{author}{\bibinfo{person}{Guorui Zhou}, \bibinfo{person}{Xiaoqiang
  Zhu}, \bibinfo{person}{Chenru Song}, \bibinfo{person}{Ying Fan},
  \bibinfo{person}{Han Zhu}, \bibinfo{person}{Xiao Ma},
  \bibinfo{person}{Yanghui Yan}, \bibinfo{person}{Junqi Jin},
  \bibinfo{person}{Han Li}, {and} \bibinfo{person}{Kun Gai}.}
  \bibinfo{year}{2018}\natexlab{}.
\newblock \showarticletitle{Deep interest network for click-through rate
  prediction}. In \bibinfo{booktitle}{\emph{Proceedings of the 24th ACM SIGKDD
  international conference on knowledge discovery \& data mining}}.
  \bibinfo{pages}{1059--1068}.
\newblock


\bibitem[Zhu et~al\mbox{.}(2021)]%
        {zhu2021learning}
\bibfield{author}{\bibinfo{person}{Yongchun Zhu}, \bibinfo{person}{Ruobing
  Xie}, \bibinfo{person}{Fuzhen Zhuang}, \bibinfo{person}{Kaikai Ge},
  \bibinfo{person}{Ying Sun}, \bibinfo{person}{Xu Zhang}, \bibinfo{person}{Leyu
  Lin}, {and} \bibinfo{person}{Juan Cao}.} \bibinfo{year}{2021}\natexlab{}.
\newblock \showarticletitle{Learning to warm up cold item embeddings for
  cold-start recommendation with meta scaling and shifting networks}. In
  \bibinfo{booktitle}{\emph{Proceedings of the 44th International ACM SIGIR
  Conference on Research and Development in Information Retrieval}}.
  \bibinfo{pages}{1167--1176}.
\newblock


\end{thebibliography}

\end{document}